\documentclass[12pt,preprint]{emulateapj}
\usepackage{graphicx}
\usepackage{epsfig}

\def\kms{km~s$^{-1}$}

\slugcomment{Submitted 2008 April 9}

\shorttitle{Late-time Optical Emission of SN 1986J}
\shortauthors{Milisavljevic et al.}
\begin{document}

\title{The Evolution of Late-time Optical Emission from SN 1986J}  

\author{Dan Milisavljevic\altaffilmark{1},
        Robert A.~Fesen\altaffilmark{1},
	Bruno Leibundgut\altaffilmark{2},
        and Robert P.~Kirshner\altaffilmark{3}}

\altaffiltext{1}{6127 Wilder Lab, Department of Physics \& Astronomy, Dartmouth
  College, Hanover, NH 03755}

\altaffiltext{2}{ESO, Karl-Schwarzschild-Strasse 2, Garching, D-85748, Germany}

\altaffiltext{3}{Harvard-Smithsonian Center for Astrophysics, 60 Garden Street,
  Cambridge, MA 02138}

\begin{abstract}

We present late-time optical images and spectra of the Type IIn supernova SN
1986J. {\sl HST} ACS/WFC images obtained in February 2003  show it to be still
relatively bright with m$_{F606W}$ = 21.4 and m$_{F814W}$ = 20.0 mag. Compared
against December 1994 {\sl HST} WFPC2 images, SN~1986J shows a decline of only
$\la$1 mag in brightness over eight years. Ground-based spectra taken in 1989,
1991 and 2007 show a 50\% decline in H$\alpha$ emission between $1989-1991$ and
an order of magnitude drop between $1991-2007$, along with the disappearance of
\ion{He}{1} line emissions during the period $1991-2007$.  The object's
[\ion{O}{1}] $\lambda\lambda$6300, 6364, [\ion{O}{2}] $\lambda\lambda$7319,
7330 and [\ion{O}{3}] $\lambda\lambda$4959, 5007 emission lines show two
prominent peaks near $-$1000 \kms \ and $-$3500 \kms, with the more blueshifted
component declining significantly in strength between 1991 and 2007.  The
observed spectral evolution suggests two different origins for SN~1986J's
late-time optical emission: dense, shock-heated circumstellar material which
gave rise to the initially bright H$\alpha$, \ion{He}{1}, and [\ion{N}{2}]
$\lambda$5755 lines, and reverse-shock heated O-rich ejecta on the facing
expanding hemisphere dominated by two large clumps generating two blueshifted
emission peaks of [\ion{O}{1}], [\ion{O}{2}], and [\ion{O}{3}] lines.

\end{abstract}


\section{Introduction}

With a peak flux density of 128 mJy at 5 GHz, SN~1986J is one of the most
radio-luminous supernovae ever detected \citep{Weiler90}. The supernova
probably occurred early in $1983$ in the edge-on galaxy NGC~891, more than
three years before its August 1986 discovery in the radio
\citep{vanGorkom86,Rupen87,Bietenholz02}. With its optical outburst going
unnoticed, the earliest optical detection showed the supernova at a magnitude
of 18.4 in $R$ in January of 1984 \citep{Rupen87,Kent87}. 

SN 1986J is classified as a Type IIn (see \citealt{Schlegel90}) and has been
compared with other luminous SNe IIn like SN 1988Z and SN 1995N. Optical
spectra of SN 1986J obtained in 1986 showed prominent and narrow ($\Delta v
\la$ 700 km s$^{-1}$) H$\alpha$ emission with no broad component
\citep{Rupen87}.  Emission lines of [\ion{O}{1}], [\ion{O}{2}], and
[\ion{O}{3}] had somewhat larger widths of 1000 $< \Delta v <$ 2000 km
s$^{-1}$.  Many narrow and weak emission lines including those from helium were
also observed. Spectra taken three years later in 1989 showed that the dominant
narrow H$\alpha$ emission had diminished in strength, with the forbidden oxygen
emission lines relatively unchanged (\citealt{Leibundgut91}; hereafter L91).     

Early very long baseline interferometry (VLBI) revealed an aspherical source
with marginal indication of an expanding shell \citep{Bartel89,Bartel91}.
Subsequent VLBI observations show a distorted shell and a current expansion
velocity of $\sim$6000 km s$^{-1}$, considerably less than an extrapolated
initial velocity of $\sim$20~000 km s$^{-1}$ \citep{Bietenholz02}.     

In this Letter, we present Hubble Space Telescope ({\sl HST}) images of SN
1986J showing it to still be relatively luminous optically more than two
decades after outburst. We also present ground-based optical spectra obtained
at three epochs spanning 18 yr to follow its late-time emission evolution.  

\section{Observations}

Images of NGC 891 in the region around SN 1986J obtained by the Advanced Camera
for Surveys (ACS) system aboard the {\sl HST} using the Wide Field Channel
(WFC) on 2003 Feb 18 and 20 were retrieved from STScI archives (GO 9414; PI:R.\
de Grijs).  The images were taken using filters F606W and F814W.  Standard
IRAF/STSDAS data reduction was done including debiasing, flat-fielding,
geometric distortion corrections, photometric calibrations, cosmic-ray and hot
pixel removal, with the STSDAS \texttt{drizzle} task used to combine exposures.  

Low-dispersion optical spectra of SN 1986J were obtained with the MMT on Mount
Hopkins using the Red Channel spectrograph with a TI $800 \times 800$ CCD on
1989 September 5 and again on 1991 October 14.  For both observations, a
2$\arcsec$ wide slit and a 150 lines mm$^{-1}$ grating was used to obtain
spectra spanning 4000--8000 \AA, with a resolution of $\sim$30 \AA. Total
exposure times for each spectrum was 9000 s.

Spectra of SN~1986J were also obtained on 2007 September 11 at MDM Observatory
using the 2.4~m Hiltner telescope with the Boller \& Chivens CCD spectrograph
(CCDS).  A north--south 1.5$\arcsec \times$ 5$\arcmin$ slit and a 150 lines
mm$^{-1}$ 4700 \AA \ blaze grating was used to obtain two sets of spectra, one
consisting of $3 \times 1800$ s exposures spanning 4000 -- 7100 \AA, and
another of $2  \times 1800$ s exposures covering 6900--9000 \AA \ using an LG
505 order separation filter.  Both spectra were of resolution of $\sim$10 \AA.
Another spectrum with this same setup consisting of $2 \times 3000$ s exposures
was taken on 2007 Dec 20 spanning 6100 -- 9000 \AA.  Seeing was around
1$\arcsec$ for all spectra.  The spectra were processed using standard
procedures in IRAF\footnote{The Image Reduction and Analysis Facility is
distributed by the National Optical Astronomy Observatories, which are operated
by the Association of Universities for Research in Astronomy, Inc., under
cooperative agreement with the National Science Foundation.} using standard
stars from \citet{Strom77}.   

\begin{figure}[hp]
\plotone{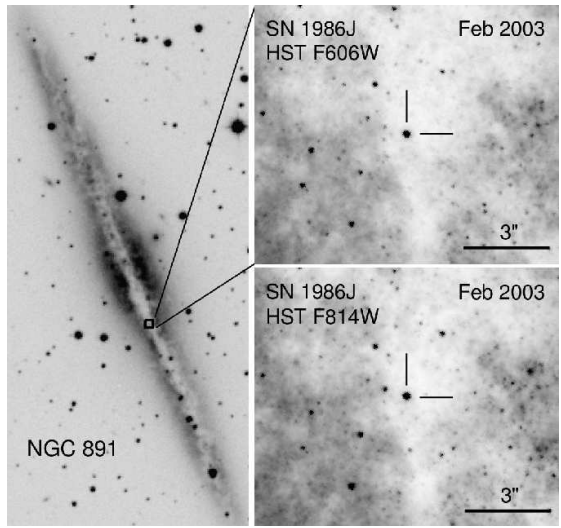}
\caption{ {\it Left}: DSS2 Blue image of NGC 891
  with the location of SN 1986J marked. {\it Right Panels}: Feb 2003 {\sl HST}
  ACS/WFC F606W and F814W images of NGC~891 centered on SN
  1986J. North is up and east is to the left.}
\label{fig:ngc891}
\end{figure}

\section{Results}

\subsection{Late-Time Optical Photometry} 

The left panel of Figure 1 shows the blue DSS2 image of NGC 891 with the
location of SN~1986J marked, and the two right panels show {\sl HST} ACS/WFC
images of NGC~891 centered on the region around the supernova.  With 2003 epoch
VEGAMAG apparent magnitudes of m$_{F606W}$ = 21.4 and m$_{F814W}$ = 20.0 mag,
these images indicate that SN~1986J has remained relatively bright nearly two
decades after the estimated 1983 optical outburst.   

The SN 1986J site in NGC~891 was also imaged some eight years earlier on 1994
December 1 by {\sl HST} with the Wide Field Planetary Camera 2 (WFPC2) using
the F606W filter (see \citealt{vanDyk99}). Reduction of these 1994 data show
m$_{F606W} =$ 21.3 mag.  Accounting for differences in instrumental response
between the WFPC2 and ACS, these observations suggest a decline of only $\la$ 1
mag in the F606W filter over the eight years separating the observations.   

\subsection{Emission Line Changes Since 1989}

In Figure 2 we present optical spectra of SN 1986J at three epochs spanning 18
years: 1989.7 (published by L91) 1991.8, and $2007$. The 2007 spectrum is an
average of the three spectra obtained in September and December of 2007, with
the combined relative fluxes believed accurate to within $\pm$20\%.

Between the three epochs, SN~1986J's optical emission shows several significant
changes.  The greatest change is the decline in H$\alpha$ emission.  As of
2007, the H$\alpha$ line observed centered around 6564 \AA \ has a flux of 4
$\times 10^{-16}$ erg s$^{-1}$ cm$^{-2}$, down some 20 times from 9 $\times
10^{-15}$ erg s$^{-1}$ cm$^{-2}$ observed in 1989.  The [\ion{N}{2}]
$\lambda\lambda$6548,6583 emission lines are not resolved, and we measure the
H$\alpha$ emission in 2007 assuming nitrogen contributes approximately 1/4 of
the total integrated flux around the line.  We estimate that H$\alpha$ emission
declined by a factor of 2 between 1989 and 1991 and by a factor of 10 between
1991 and 2007.

The line center for SN~1986J's H$\alpha$ emission also showed an increasing
blueshift between 1989 and 2007.  In 1986, the line center roughly matched
NGC~891's redshift of +528 km s$^{-1}$ \citep{Rupen87,deVaucouleurs91}.
However, by 1989 the shift had decreased to +330 km s$^{-1}$ (L91), and in 2007
the center of the observed H$\alpha$ emission shifted still more to the blue,
virtually negating the galaxy's systemic velocity and appearing practically
unredshifted. 
 
Other changes in the late-time spectra of SN~1986J include the fading beyond
detectability of H$\beta$ and He I $\lambda$5876 and $\lambda$7065 emission
lines in the 2007 spectrum which were present in 1989 and 1991. Also
considerably diminished in 2007 is the emission associated with the
[\ion{O}{3}] $ \lambda\lambda$4959, 5007 emission observed around 4980 \AA.  

Changes in the profiles of some emission lines are evident.  Broad emission
around 7300 \AA \ consisting of two prominent emission peaks at $\simeq$7250
\AA \ and 7320 \AA \ seen in both the 1989 and 1991 spectra exhibits a
significant diminishment along its blue side in the 2007 spectrum (see Fig.\
2). We identify both emission peaks with the [\ion{O}{2}] $\lambda\lambda$7319,
7330 line emission. Although some contribution from [\ion{Ca}{2}]
$\lambda\lambda$7291,7324 is possible, [\ion{O}{2}] emission likely dominates
the broad $7200-7350$ \AA \ emission.

The bluer emission peak at 7250 \AA, prominently visible in 1989 and 1991
spectra, faded significantly by 2007, evolving into a weak, broad emission
feature blueward of 7300 \AA.  The other emission peak at 7320 \AA \ showed a
smaller intensity decline and a shift to the red by $\sim$10 \AA \ relative to
its appearance in 1989.  Lastly, faint redshifted [\ion{O}{2}] emission
extending from 7380 to 7450 \AA \ visible in the 1989 spectrum gradually
weakened and decreased in velocity in the 1991 and 2007 spectra.

Other features show only minor changes in strength and/or profile.  The broad
emission centered around 6295 \AA \ identified with the [\ion{O}{1}]
$\lambda$6300 line has a 2007.8 epoch flux of 1.4 $\times 10^{-15}$ erg
s$^{-1}$ cm$^{-2}$ assuming a ratio of 3:1 for the 6300 \AA \ and 6364 \AA \
lines.  This is roughly the same as the combined flux of 2.7 $\times 10^{-15}$
erg s$^{-1}$ cm$^{-2}$ for the two lines lines measured in 1989 by L91.  

\section{Discussion}

A steady drop in H$\alpha$ emission strength together with smaller declines in
forbidden oxygen emissions are consistent with our estimated m$_{F606W} \la$ 1
mag decline between 1994 and 2003.  To investigate possible changes in emission
since then, we used \texttt{synphot} to compare the count rate of the 2003
F606W image against the expected count rate of the ACS/WFC F606W given the 2007
spectrum as input.  The rates are marginally different and within the
uncertainties associated with the relative flux of the spectra and light loss
from the slit.  This suggests that SN~1986J's optical flux has not deviated
appreciably from a continued, slow decline over the last four years.   

\begin{figure*}[htp]
\begin{center}
\epsscale{0.95}
\plotone{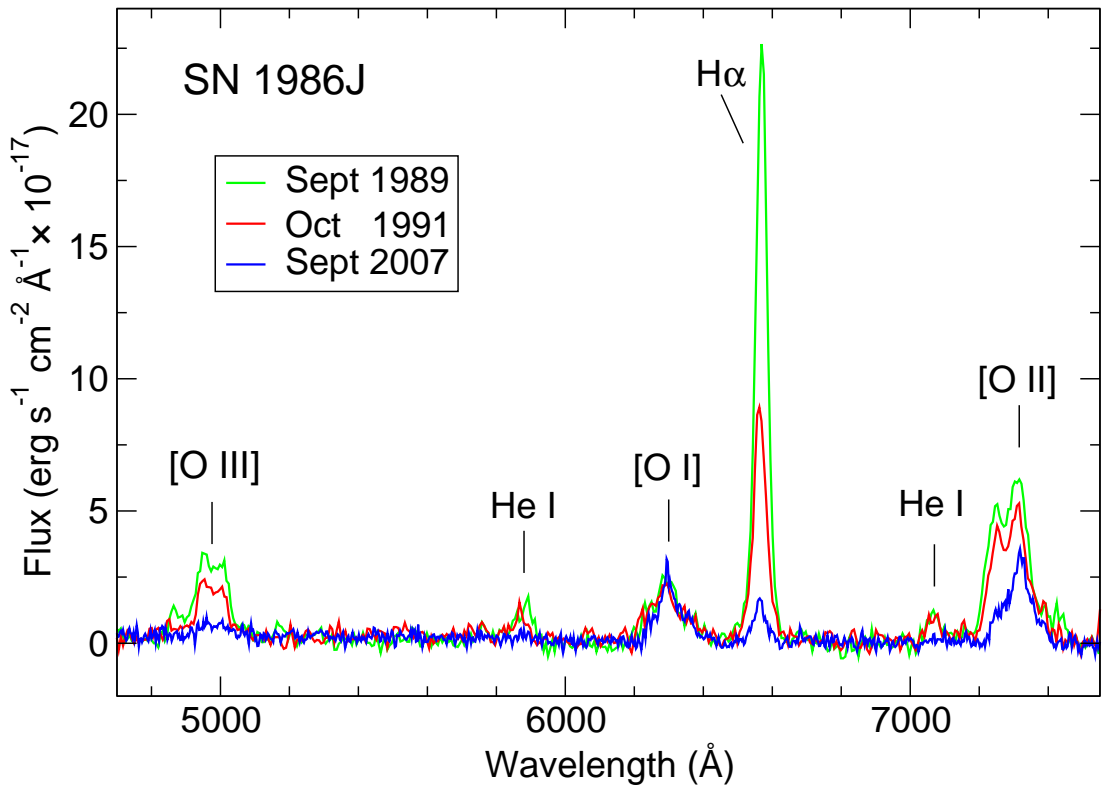}
\caption{Optical spectra of SN 1986J at three epochs covering 18 years.} 
\label{fig:spectrum}
\end{center}
\end{figure*}

\newpage

\subsection{Emission-Line Profiles from O-Rich SN Ejecta}

While a blueshifted, double-peak emission profile is most apparent in the
[\ion{O}{2}] $\lambda\lambda$7319,7330 lines in the 1989 and 1991 spectra, in
fact, all of SN~1986J's oxygen emissions display a similar double-peak emission
profile blueshifted with respect to the host galaxy's rest frame. Figure 3
presents an overlay of [\ion{O}{1}], [\ion{O}{2}], and [\ion{O}{3}] line
profiles plotted in velocity space. This figure shows good agreement for both
the line profiles and emission peaks near $-$1000 and $-$3500 \kms.   

Added support for double-peak oxygen line profiles comes from faint emission
near 4340 \AA \ present in both 1989 and 1991 spectra which we interpret as
[\ion{O}{3}] $\lambda$4363 line emission (see Fig.\ $2$ in L91).  When
corrected for NGC~891's redshift, the positions and widths of the peaks
observed at 4313 \AA \ and 4338 \AA \ match the $-$1000  \kms \ and $-$3500
\kms \ emission peaks observed in the other oxygen profiles. After correcting
for foreground reddening of $A_{V} = 1.5$ mag \citep{Rupen87}, the observed
[\ion{O}{3}] I(4959+5007)/I(4363) line ratio $\simeq2$ suggests an electron
density for the [\ion{O}{3}] emitting region of $(3-5) \times 10^{6}$ cm$^{-3}$
assuming an electron temperature of $(2.5-5.0) \times 10^4$ K like that found
in shock-heated O-rich ejecta seen in young supernova remnants \citep{HF96,Blair00}.

An interpretation of a spectrum dominated by two blueshifted, O-rich ejecta
clumps is a very different one from that proposed by L91 for explaining the
box-like [\ion{O}{3}] $\lambda\lambda$4959, 5007 profile. They suggested that
the observed shape was due to the $\lambda$5007/$\lambda$4959 line ratio being
close to 1:1 instead of the optically thin ratio of 3:1 typically observed in
low density nebulae. A ratio close to unity for both the [\ion{O}{1}] and
[\ion{O}{3}] line doublets were interpreted as caused by emission originating
from regions with electron densities of $n_{e} \sim 10^{9}$ cm$^{-3}$. However,
in light of the strong similarity of all oxygen emission profiles, such high
density estimates appear no longer valid.

\subsection{Origin of the Late-Time Optical Emission}

Our interpretation of line emission profiles together with the observed
spectral evolution over the last two decades suggests two separate sites for
SN~1986J's late-time optical emission. The decline of SN 1986J's H$\alpha$
emission and its relatively low expansion velocity ($< 700$ km s$^{-1}$)
suggests this emission comes from shock-heated circumstellar material (CSM).
Early spectra showing an initially very bright H$\alpha$ emission along with
fainter emissions from \ion{He}{1} and [\ion{Fe}{2}] are consistent with an
emission nebula generated by a $\sim 1.5 \times 10^{4}$ \kms \ blast wave
overrunning a dense CSM environment rich in CNO-processed material
\citep{Rupen87}. The apparent blueshift in the line center of H$\alpha$ over
the past 20+ years is likely due to increasing extinction of the receding
hemisphere possibly due to dust formation in the SN ejecta.     

\citet{Chugai94} suggested SN~1986J's late-time optical emission originates
from shocked dense clouds of circumstellar gas in the progenitor star's clumpy
pre-SN wind. The presence of [\ion{N}{2}] $\lambda$5755 line emission and the
lack of strong [\ion{N}{2}] $\lambda\lambda$6548, 6583 emission \citep{Rupen87}
supports this scenario, suggesting relatively high densities ($n_{e}$ $\sim
10^{6}$ cm$^{-3}$) similar to that seen in the circumstellar ring around
SN~1987A.   

\begin{figure}[htp]
\begin{center}
\includegraphics[width=7.5cm,clip=true]{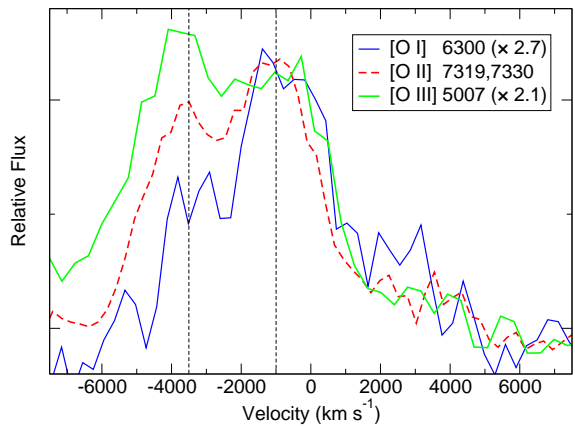}
\caption{SN 1986J's 1989 oxygen line profiles.
  Velocities shown are with respect to 6300, 7325, and 5007 \AA \ in the
rest frame of NGC 891. Vertical dashed lines are positioned at $-$1000 km
s$^{-1}$ and $-$3500 km s$^{-1}$.} 
\end{center}
\end{figure}

The interaction of the SN's outward-moving blast wave with dense and clumpy CSM
will generate a strong reverse-shock into slower expanding SN ejecta, leading
to the observed forbidden oxygen line emissions.  The presence of two prominent
blueshifted emission peaks across three ionization stages implies this
component of SN~1986J's optical emission comes mainly from two large patches of
reverse shock-heated, O-rich ejecta on the facing expanding hemisphere having
radial velocity components in our line-of-sight around $-1000$ \kms \ and
$-3500$ \kms.

The gradual redward shift of the $-1000$ \kms \ emission component toward
smaller blueshifted velocities seen most clearly in the [\ion{O}{2}]
$\lambda\lambda$7319,7330 profile between 1989 and 2007 may signal the
progression of reverse shock emission coming from inner, slower moving O-rich
ejecta during the intervening two decades. Additionally, weak emission seen
redward of 7330 \AA \ together with weak emission near 5050 \AA \ possibly
associated with [\ion{O}{3}] might indicate highly reddened O-rich ejecta
located on the rear expanding expanding hemisphere with radial velocities up to
$3500$ \kms.

Finally, mention should be made concerning the possibility for photoionization
of SN~1986J's ejecta by its bright central compact source \citep{Chevalier87}.
Early optical and radio observations of SN 1986J suggested that it may be a
very young Crab Nebula-like remnant \citep{Chevalier87,Weiler90} and such a
connection has been strengthened by recent VLBI observations showing a bright,
compact radio component with an inverted spectrum near the center of the
expanding shell \citep{Bietenholz04,Bietenholz08}. This central source is
thought to be either emission from a young, energetic neutron star or accretion
onto a black hole.  The optical filaments in the Crab Nebula are mainly
photoionized by its pulsar. With SN 1986J’s central component some 200 times
more luminous than the Crab Nebula between 14 and 43 GHz \cite{Bietenholz08},
this raises the possibility of photoexcitation of SN~1986J's ejecta. 

However, during its first decade of evolution, SN 1986J's strong [\ion{O}{3}]
$\lambda$4363 line suggested temperatures more indicative of shock heating ($T
\ga 25 000$ K) rather than photoionization ($T \leq 15 000$ K).  Additionally,
the high densities of the O-rich ejecta and/or formation of dust in the ejecta
could limit the importance of photoionization by the central source. Indeed,
the large Balmer decrement ratio of H$\alpha$/H$\beta$ $\sim 45$ (1986.8 epoch;
L91) observed in early spectra may be an indication of high internal
extinction.  

At its current age ($\sim25$ yr), the importance of photoexcitation from the
central source, quite possibly a young Crab-like neutron star, is less clear.
Recent observations with {\sl XMM-Newton} and {\sl Chandra} have shown a sharp
decline in X-ray luminosity, perhaps signaling a diminishing role of
shock-heating in SN 1986J's late-time optical emission \citep{Temple05}.
Future increased contribution from photoionization could be reflected as
broadening in optical emission line widths like that predicted by
\citet{Chevalier94}. 

In view of SN~1986J's strong oxygen line emissions, a better comparison than
the Crab might be the $\simeq 1000$ yr old LMC remnant 0540-69.3.  This remnant
has a bright pulsar wind nebula surrounding a 50 ms pulsar and shock-heated,
O-rich ejecta expanding at velocities $\sim2000$ km s$^{-1}$ \citep{Morse06}
much more in line with what is observed in SN~1986J.  

\acknowledgements

We thank R. Chevalier for helpful comments on an earlier draft.
This research was supported in part by a Canadian NSERC award to DM.

\end{document}